\documentclass[12pt]{iopart}
\usepackage{epsf}
\begin{document}

\title{On the rate of $T_{c}$ suppression by the interband impurity scattering in MgB$_{2}$} 

\author{Bo\v zidar Mitrovi\' c
\footnote[3]{To
whom correspondence should be addressed (mitrovic@brocku.ca)}}

\address{Physics Department, Brock University, St. Catharines, Ontario,
Canada L2S~3A1}

\begin{abstract}
We calculate the change in the superconducting transition temperature $T_{c}$ of 
MgB$_{2}$ caused by interband nonmagnetic impurity scattering using the Eliashberg 
theory for the two-band model of this compound. Much slower rate of $T_{c}$ suppression 
is obtained compared to the prediction based on the BCS treatment of the two-band model 
which ignores renormalization and damping associated 
with the electron-phonon interaction. Hence, the interband
impurity scattering rates deduced from experiments on MgB$_{2}$ using the  
formula which results from the BCS approach to the two-band model are underestimated. 
We generalize the BCS treatment of the two-band model to include renormalization 
effects of the electron-phonon interaction and find an excellent agreement with the 
full strong coupling calculation.

\end{abstract}



\maketitle

\section{Introduction}
There is a large body of experimental [1-14] and theoretical (for a review see [15]) evidence that MgB$_{2}$ is 
a multiband superconductor which is well described by an effective two-band model [16].
In the case of a multiband superconductor one expects that the superconducting transition 
temperature $T_{c}$ is reduced by the interband nonmagnetic (i.e.~normal) impurity scattering in 
analogy to the effect of such scattering on anisotropic single band superconductors [17,18]. 
Several years before the discovery of superconductivity in MgB$_{2}$ 
the problem of impurity scattering in a multiband 
superconductor was examined in detail by Golubov and Mazin [19] using the weak coupling BCS-type 
treatment of the pairing interaction. They obtained an equation for the change in $T_{c}$ with 
the interband impurity scattering rate which is analogous to the Abrikosov-Gor'kov formula for 
the  $T_{c}$-suppression by paramagnetic impurity scattering in ordinary superconductors. 
The BCS-type treatment of Ref.~[19] predicts that the $T_{c}$ is reduced by about 40 \% for
the interband scattering rate comparable to $k_{B}T_{c}$. For MgB$_{2}$ that would imply a drop 
in $T_{c}$ from 39 K to about 25 K for the interband impurity scattering rate $\Gamma \equiv1/(2\tau)
\equiv\gamma/2$ of about 1.7 meV. Thus, it was thought that observation of $T_{c}$ suppression with 
increasing disorder would provide the final evidence for the two-band model of MgB$_{2}$.

Experimentally, however, the situation appears to be more complicated. On one hand, as pointed out in [20],
the transition temperatures of different samples of MgB$_{2}$ are rather insensitive to 
their respective residual resistivities : the $T_{c}$s of samples with residual resistivities
in the range from 0.4 to 30 $\mu\Omega$cm differ by at most 5\%. On the other hand, irradiation of 
a polycrystalline sample of MgB$_{2}$ by fast neutrons led to an increase of residual resistivity and
reduction in $T_{c}$ by as much as 20\% [13]. 
The apparent lack of correlation between $T_{c}$ and residual resistivity in unirradiated samples
was explained [20] by very small 
values of the interband impurity scattering matrix elements because of the particular electronic
structure of MgB$_{2}$ so that the DC transport in this compound at low temperatures is 
primarily determined by {\em intraband} scattering which does not affect $T_{c}$ [19],  
while very weak interband scattering
leads to no significant change in $T_{c}$. The arguments in [20] apply to common 
substitutional impurities in MgB$_{2}$ which do not distort the lattice and subsequently Erwin 
and Mazin [21] proposed that substituting Mg with Al and/or Na would produce lattice distortions that
could lead to large enough interband impurity scattering rates to cause a reduction of 
$T_{c}$ by a couple of degrees as predicted theoretically in [19]. Presumably the irradiation by fast 
neutrons generates enough lattice distortions to cause a 20\% drop in $T_{c}$ [13].

Nevertheless, the break junction tunneling experiments on MgB$_{2}$ [4] clearly indicate that the interband
impurity scattering is significant even in undoped and unirradiated samples.
Namely, the only justification for using the equations of McMillan tunneling model for proximity effect
[22] in analyzing the break junction data on 
MgB$_{2}$ is provided by the work of Schopohl and Scharnberg [23] on tunneling density of states of a 
disordered two-band superconductor. The fact that in the latter case the equations have the 
form identical to those of the McMillan tunneling model for proximity effect is a pure accident as 
is evident from the entirely different meaning of the quasiparticle scattering rates in the two cases.
The interband scattering rates used
to fit the tunneling data [4] were at least as large as those predicted for Al/Na doped MgB$_{2}$  
($\Gamma$s were in the range from 1 to 4 meV), 
but the $T_{c}$ of the material was reported to be 39 K - close to the maximum value for 
MgB$_{2}$ of 39.4 K. A possible solution to this contradiction is that the weak coupling BCS-type 
treatment of impurity scattering in a multiband superconductor used in [19] is not {\em quantitatively}
accurate for MgB$_{2}$. The calculated electron-phonon interactions in MgB$_{2}$ [16] indicate
that it is a medium-to-strong coupling superconductor (the largest calculated electron-phonon
parameter $\lambda_{\sigma\sigma}$ for $\sigma$-band electrons is comparable to the one in Nb) and
renormalization and damping effects could play an important role in determining the rate of 
$T_{c}$ suppression by interband impurity scattering. 

In section 2 we solve the Eliashberg equations for a two-band superconductor with 
nonmagnetic impurity scattering and calculate the transition temperature as a function of 
the impurity scattering rate using realistic interaction parameters for MgB$_{2}$ [16,24].  
We find that the $T_{c}$ is suppressed by interband scattering at much slower rate than what 
was obtained using the BCS treatment in Ref.~[19]. In the same section we present the functional 
derivatives $\delta T_{c}/\delta\alpha^2F_{ij}$, $i,j=\sigma,\pi$ [25] for several representative 
impurity interband scattering rates which show how the sensitivity of $T_{c}$ to various 
electron-phonon couplings changes with impurity scattering. 
In section 3 we generalize the
BCS approach to include the renormalization caused by the electron-phonon interaction by extending
the well known $\theta$-$\theta$ model [26] to the two-band case. The numerical solution of such a 
model is found to be in excellent agreement with the full strong coupling calculation. In 
section 4 we give a summary.
\section{Strong coupling calculation}
\subsection{Formalism}
The Eliashberg equations for $T_{c}$ of a superconductor with several isotropic bands
$i=1,2,\dots$ which include nonmagnetic impurity 
scattering described by the Born approximation are [25,26]   
\begin{equation}
\phi_{i}(n)  =  \phi_{i}^{0}(n)+\sum_{j}\frac{1}{2\tau_{ij}}\frac{\phi_{j}(n)}{|\omega_{n}|Z_{j}(n)}\>,
\end{equation}
\begin{equation}
\phi_{i}^{0}(n)  =  \pi T_{c}\sum_{j}\sum_{m=-\infty}^{+\infty}[\lambda_{ij}(n-m)-\mu_{ij}\theta(E_{F}-|\omega_{m}|)]
\frac{\phi_{j}(m)}{|\omega_{m}|Z_{j}(m)}\>,
\end{equation}
\begin{equation}
\omega_{n}Z_{i}(n)  =  \omega_{n}Z_{i}^{0}(n)+\sum_{j}\frac{1}{2\tau_{ij}}\frac{\omega_{n}}{|\omega_{n}|}\>,
\end{equation}
\begin{equation}
\omega_{n}Z_{i}^{0}(n)  =  \omega_{n}+\pi T_{c}\sum_{j}\sum_{m=-\infty}^{+\infty}
\lambda_{ij}(n-m)\frac{\omega_{m}}{|\omega_{m}|}\>.
\end{equation}
Here $\phi_{i}(n)$ is the pairing self-energy in the band $i$ at the Matsubara frequency 
$\omega_{n}=\pi T_{c}(2n-1)$ and $Z_{i}(n)$ is the corresponding renormalization function 
(for a review of the Eliashberg theory of superconducting $T_{c}$ see [26]). The part 
$\phi_{i}^{0}(n)$, Eq.~(2), results from the intrabend and interband electron-phonon and 
screened Coulomb interactions, while the second term in Eq.~(1) represents the impurity 
scattering contribution. In the same way, $Z_{i}^{0}(n)$ is the contribution to the 
renormalization function from the intraband and interband electron-phonon interaction and 
the second term in Eq.~(3) gives the impurity contribution to $Z_{i}(n)$. 
The cutoff $E_{F}$ on the sums over the Matsubara frequencies $\omega_{m}$ in Eq.~(2) is  
initially taken to be large enough so that the Coulomb repulsion parameters are given by 
$\mu_{ij}=V_{ij}^{c} N_{j}$, where $V_{ij}^{c}$ is the Fermi surface averaged screened Coulomb matrix 
element between the states in the bands $i$ and $j$ ($V_{ij}^{c}=V_{ji}^{c}$) and $N_{j}$ is the Fermi 
surface density of states in band $j$ [25]. The electron-phonon coupling functions $\alpha^{2}F_{ij}(\Omega)=
\alpha^{2}f_{ij}(\Omega)N_{j}$ ($\alpha^{2}f_{ij}(\Omega)=\alpha^{2}f_{ji}(\Omega)$) enter via parameters 
$\lambda_{ij}(n-m)$
\begin{equation}
\lambda_{ij}(n-m)=\int_{0}^{+\infty}d\Omega\alpha^{2}F_{ij}(\Omega)\frac{2\Omega}{\Omega^{2}+
(\omega_{n}-\omega_{m})^{2}}\>,
\end{equation}
and the impurity scattering rates $\gamma_{ij}\equiv 1/\tau_{ij}$ are given by (we use the units in which 
$\hbar$=1 and the Boltzmann's constant $k_{B}$=1)
\begin{equation}
\frac{1}{2\tau_{ij}}=n_{imp}\pi N_{j}|V_{ij}|^{2}
\end{equation}
where $n_{imp}$ is the concentration of nonmagnetic impurities and $V_{ij}$ is the Fermi surface averaged 
matrix element of the change in the lattice potential caused by an impurity between the states 
in the bands $i$ and $j$. Clearly, $\gamma_{ij}/\gamma_{ik}=N_{j}/N_{k}=\lambda_{ij}/\lambda_{ik}$, where 
$\lambda_{ij}=\lambda_{ij}(0)$ (see equation (5)). 

In principle, Eqs.~(1)-(4) have the form of an 
eigenvalue problem of a temperature dependent matrix with eigenvector $\hat{\phi}$, and $T_{c}$ is 
determined as the heighest temperature at which the largest eigenvalue of the matrix is one. However,  
before such solution is attempted one can simplify the problem further. First, by introducing the 
gap function as renormalized pairing self-energy $\Delta_{i}(n)=\phi_{i}(n)/Z_{i}(n)$ one can eliminate the 
{\em intraband} impurity scattering from the problem by combining Eqs.~(1) and (3)
\begin{equation}
\Delta_{i}(n)\left(Z_{i}^{0}(n)+\sum_{j\neq i}\frac{1}{2\tau_{ij}}\frac{1}{|\omega_{n}|}\right)= 
 \phi_{i}^{0}(n)+\sum_{j\neq i}\frac{1}{2\tau_{ij}}\frac{\Delta_{j}(n)}{|\omega_{n}|}\>,
\end{equation}
where $\phi_{i}^{0}(n)$ is given by Eq.~(2) with $\phi_{j}(m)/Z_{j}(m)$ replaced by $\Delta_{j}(m)$. 
Next, the eigenvalue problem can be symmetrized by defining 
\begin{equation}
u_{i}(n)=\sqrt{N_{i}}\frac{\Delta_{i}(n)}{|\omega_{n}|}\sqrt{|\omega_{n}|Z_{i}^{0}(n)+
\sum_{j\neq i}\frac{1}{2\tau_{ij}}}\>
\end{equation}
together with $\lambda_{ij}^{s}(n-m)=\sqrt{N_{i}/N_{j}}\lambda_{ij}(n-m)$, 
$\mu_{ij}^{s}=\sqrt{N_{i}/N_{j}}\mu_{ij}$ and $1/(2\tau_{ij}^{s})=\sqrt{N_{i}/N_{j}}/(2\tau_{ij})$.
With these definitions Eq.~(7) reduces to 
\begin{equation}
\fl u_{i}(n)=\varepsilon(T)\sum_{j}\sum_{m=-\infty}^{\infty}\pi T\frac{\lambda_{ij}^{s}(n-m)-\mu_{ij}^{s}
\theta(E_{F}-|\omega_{m}|)+\frac{1}{2\pi T\tau_{ij}^{s}}(1-\delta_{ij})\delta_{nm}}
{\sqrt{|\omega_{n}|Z_{i}'(n)}\sqrt{|\omega_{m}|Z_{j}'(m)}}u_{j}(m)\>,
\end{equation}
with
\begin{equation}
Z_{i}'(n)=Z_{i}^{0}(n)+\sum_{j\neq i}\frac{1}{2\tau_{ij}}\frac{1}{|\omega_{n}|} 
\end{equation}
and $\varepsilon(T)$=1 when $T=T_{c}$. Finally, the size of the matrix which has do be diagonalized 
can be reduced by cutting off
the Matsubara sums in (9) at a smaller energy $\omega_{c}$ which is still large enough so that
$Z_{i}'(n)\approx$1 for $|\omega_{n}|>\omega_{c}$; hence, $\omega_{c}$ has to be at least 5-10 times 
the maximum phonon energy $\Omega_{m}$ in various spectral functions $\alpha^{2}F_{ij}(\Omega)$ and much
larger than the largest band off-diagonal $1/2\tau_{ij}$. The reduction in cutoff from $E_{F}$ to $\omega_{c}$ is 
accompanied by replacement of the Coulomb repulsion parameters  $\mu_{ij}^{s}$ in (9) with $\mu_{ij}^{*}
(\omega_{c})$ where the matrix (in band indices) $\hat{\mu}^{*}(\omega_{c})$ is related to 
matrix $\hat{\mu}^{s}$ by [25]
\begin{equation}
\hat{\mu}^{*}(\omega_{c})=\left(\hat{1}+\hat{\mu}^{s}\ln\frac{E_{F}}{\omega_{c}}
\right)^{-1}\hat{\mu}^{s}\>.
\end{equation}

\subsection{Numerical Results}
We solved Eqs.~(9)-(11) using the spectral functions $\alpha^{2}F_{\sigma\sigma}$, $\alpha^{2}F_{\sigma\pi}$, 
$\alpha^{2}F_{\pi\pi}$ and $\alpha^{2}F_{\pi\sigma}$ for MgB$_{2}$
obtained from the first principle electronic structure calculations and presented in [16].
The corresponding coupling parameters given by Eq.~(5) with $\omega_{n}-\omega_{m}$ = 0 are  
$\lambda_{\sigma\sigma}$ = 1.017, $\lambda_{\sigma\pi}$ = 0.212, $\lambda_{\pi\pi}$ = 0.446 and 
$\lambda_{\pi\sigma}$ = 0.155. Since $\alpha^{2}F_{ij}(\Omega)=\alpha^{2}f_{ij}(\Omega)N_{j}$, 
with $\alpha^{2}f_{ij}(\Omega)=\alpha^{2}f_{ji}(\Omega)$, these values of $\lambda$-parameters 
fix the ratio of the partial band densities of states 
$N_{\pi}/N_{\sigma}$ = $\lambda_{\sigma\pi}/\lambda_{\pi\sigma}$ at 1.37. That fixes the ratio $\gamma_{\sigma\pi}
/\gamma_{\pi\sigma}$ (see Eq.~(6)) and we chose $\gamma_{\pi\sigma}$ as the independent scattering parameter. 

To minimize the effect of changes in the number $N_c=[\omega_{c}/(2\pi T_c) +0.5]$ of Matsubara 
frequencies ($[\cdots]$ denotes the integer part) on our numerical results as $T_c$ is reduced by increased 
interband impurity scattering rate we had to take the cutoff $\omega_{c}$ to be at least 10 times the 
maximum phonon energy $\Omega_{m}$. With $\omega_{c}$ fixed at 1000 meV the Coulomb repulsion 
parameters $\mu_{\sigma\sigma}^{*}$, $\mu_{\sigma\pi}^{*}$, $\mu_{\pi\pi}^{*}$, $\mu_{\pi\sigma}^{*}$ 
were determined as follows. Choi {\em et al.} [24] calculated the ratios of the {\em screened}  
Coulomb repulsion parameters for MgB$_{2}$ to be
$\mu_{\sigma\sigma}$ : $\mu_{\pi\pi}$ : $\mu_{\sigma\pi}$ : $\mu_{\pi\sigma}$ = 1.75 : 2.04 : 1.61 : 1.00. 
Since $\mu_{ij}=V_{ij}^{c} N_{j}$, the ratio $\mu_{\sigma\pi}/\mu_{\pi\sigma}$ = 1.61  
implies that in their calculation $N_{\pi}/N_{\sigma}$ = 1.61, which is considerably higher 
than the value of
1.37 found in [16] and adopted by us in this work by our choice of the electron-phonon coupling spectra. 
Leaving aside the reasons for such a discrepancy in $N_{\pi}/N_{\sigma}$ between the two sets of 
electronic structure calculations we use the ratios of the $\mu$-values calculated in [24]  
to extract from them the 
ratios of the {\em screened} Coulomb matrix elements: 
$V_{\sigma\sigma}^{c}$ : $V_{\pi\sigma}^{c}$ = 1.75 : 1.00, $V_{\pi\pi}^{c}$ : $V_{\sigma\pi}$ = 2.04 : 1.61 
and, because $V_{\pi\sigma}^{c}$ = $V_{\sigma\pi}^{c}$, $V_{\sigma\sigma}^{c}$ : $V_{\pi\pi}^{c}$ = 1.75 : 1.267. 
These could be combined with $N_{\pi}/N_{\sigma}$ = 1.37 to produce the ratios $\mu_{\sigma\sigma}/
\mu_{\pi\pi}$ = 1.01, $\mu_{\sigma\sigma}/\mu_{\sigma\pi}$ = 1.28 and $\mu_{\sigma\sigma}/\mu_{\pi\sigma}$ = 
1.75 leaving the single fitting parameter $\mu_{\sigma\sigma}$ once a choice is made for the initial cutoff 
$E_F$ (see Eq.~(11)).  
Since $E_{F}$ is on the order of the largest electronic energy scale in the problem, 
we took $E_{F}$ to be equal to the $\pi$-bandwidth of 15 eV [27] and
fitted $\mu_{\sigma\sigma}$ in the four equations implied by the 2$\times$2 matrix equation (11)
\begin{eqnarray}
\mu_{\sigma\sigma}^{*}(\omega_{c}) & = &  \left[\mu_{\sigma\sigma}+
(\mu_{\sigma\sigma}\mu_{\pi\pi}-\mu_{\sigma\pi}\mu_{\pi\sigma})
\ln\frac{E_{F}}{\omega_{c}}\right]/D\>, \\
\mu_{\sigma\pi}^{*}(\omega_{c}) & = & \sqrt{N_{\sigma}/N_{\pi}}\mu_{\sigma\pi}/D\>, \\
\mu_{\pi\pi}^{*}(\omega_{c}) & = &  \left[\mu_{\pi\pi}+
(\mu_{\sigma\sigma}\mu_{\pi\pi}-\mu_{\sigma\pi}\mu_{\pi\sigma})
\ln\frac{E_{F}}{\omega_{c}}\right]/D\>, \\
\end{eqnarray}
\begin{equation}
D  =   1+(\mu_{\sigma\sigma}+\mu_{\pi\pi})\ln\frac{E_{F}}{\omega_{c}}
 (\mu_{\sigma\sigma}\mu_{\pi\pi}-\mu_{\sigma\pi}\mu_{\pi\sigma})
\left(\ln\frac{E_{F}}{\omega_{c}}\right)^{2}\>,
\end{equation} 
to the experimental $T_{c0}$ of 39.4 K for the case of no impurity scattering. The results were 
$\mu_{\sigma\sigma}$ = 0.848234 with $\mu_{\sigma\sigma}^{*}(\omega_{c})$ = 0.225995, 
$\mu_{\pi\pi}^{*}(\omega_{c})$ = 0.225010 and $\mu_{\sigma\pi}^{*}(\omega_{c})=
\mu_{\pi\sigma}^{*}(\omega_{c})$ = 0.067148. 

In Figure 1 we show with the solid line the calculated $T_c/T_{c0}$ as a function of $\gamma_{\pi\sigma}/T_{c0}$ 
(note that for $\gamma_{\pi\sigma}/T_{c0}$ $\geq$ 2 the scale is logarithmic).
The dotted line represents the prediction based on the BCS weak coupling approach of Ref.~[19] where  $T_c$ 
drops initially with the slope $-\pi/8$ (see Fig.~1 and Eq.~(13) in [19]). Clearly the full strong coupling 
calculation with realistic electron-phonon spectral functions and Coulomb repulsion parameters leads to a 
much slower drop in $T_c$ with increasing interband impurity scattering rate than what was obtained in [19] 
using the BCS approach: for $\gamma_{\pi\sigma}$ = $T_{c0}$ we get a drop in transition temperature of only 
8 \%, while the BCS treatment predicts a drop of about 36 \%.  

\begin{figure}
\begin{center}
\epsfbox{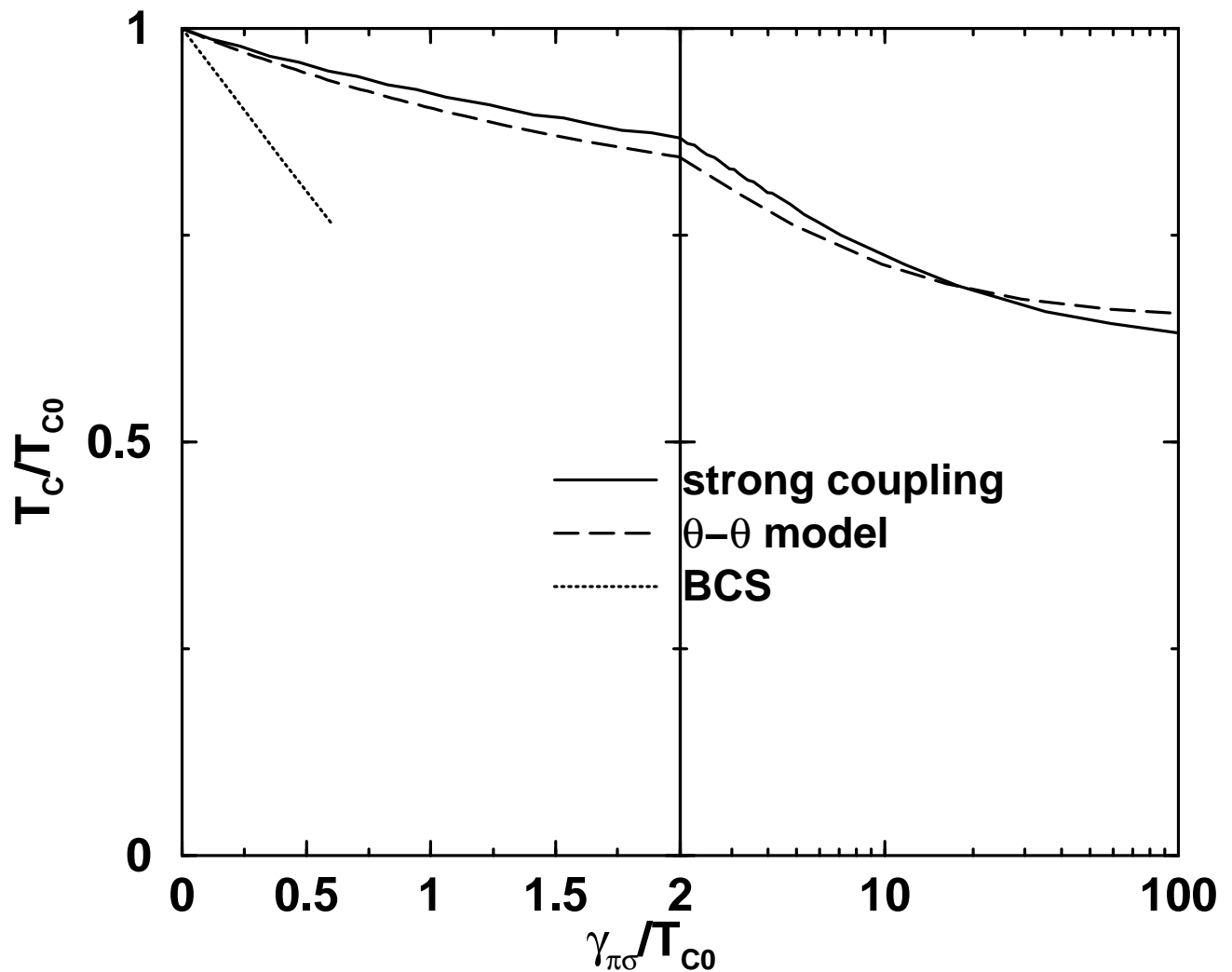}
\end{center}
\caption{\label{label}The relative change in the transition temperature $T_{c}/T_{c0}$ as a function 
of the interband impurity scattering rate parameter $\gamma_{\pi\sigma}/T_{c0}$. The solid curve 
gives the results of the full strong coupling calculation for MgB$_{2}$ and the long dashed curve 
gives the results obtained from $\theta$-$\theta$ model with interaction parameters that were used 
in the strong coupling calculation. The dotted line indicates the weak coupling result from Ref.~[19] 
which predicts for low values of $\gamma_{\pi\sigma}$ a straight line with the slope 
$-\pi/8$.
}
\end{figure}
 
Before we address in the next section 
the reasons for such a large discrepancy between the strong coupling and the BCS results
we give in Figs.~(2) and (3) calculated functional derivatives
$\delta T_{c}/\delta\alpha^2F_{ij}$, $i,j=\sigma,\pi$ which show how the sensitivity of $T_{c}$ 
to various electron-phonon couplings changes with increasing 
interband impurity scattering. In [25] these functional derivatives were computed for the case of no 
impurity scattering and it was found that the band-diagonal functional derivatives 
$\delta T_{c}/\delta\alpha^2F_{\sigma\sigma}$ and $\delta T_{c}/\delta\alpha^2F_{\pi\pi}$  
have similar shapes but their overall magnitudes differ due to difference  
in sizes of the gap-functions $\Delta_{\sigma}(n)$ and $\Delta_{\pi}(n)$ in the two bands. 
This is also the case for the lowest $\gamma_{\pi\sigma}$ in Fig.~2, but as the scattering 
rate increases the difference in magnitudes of the two gaps becomes smaller and the overall 
magnitudes of the two band-diagonal functional derivatives become comparable. For $\gamma_{\pi\sigma} $ =
100$T_{c0}$ the magnitude of $\delta T_{c}/\delta\alpha^2F_{\pi\pi}$ is larger than the magnitude of 
$\delta T_{c}/\delta\alpha^2F_{\sigma\sigma}$ presumably because $\lambda_{\pi\pi}<\lambda_{\sigma\sigma}$ 
[28] and the difference in sizes of the gaps $\Delta_{\sigma}(n)$ and $\Delta_{\pi}(n)$ 
has largely disappeared. Another consequence of this disappearance of the difference between 
the gaps in the two bands is that with increasing $\gamma_{\pi\sigma}$ the shapes 
of $\delta T_{c}/\delta\alpha^2F_{\pi\sigma}$ and $\delta T_{c}/\delta\alpha^2F_{\sigma\pi}$, Fig.~3,
become more and more 
similar to the shapes of the band-diagonal functional derivatives. The divergencies at $\Omega$ = 0 still 
persist, but are progressively confined to smaller and smaller neighborhoods of $\Omega$ = 0. Again, the local 
maximum in $\delta T_{c}/\delta\alpha^2F_{\pi\sigma}$ near $\Omega$ = 8$k_{B}T_{c}$ for $\gamma_{\pi\sigma}$ =
100$T_{c0}$ is higher than the corresponding maximum in $\delta T_{c}/\delta\alpha^2F_{\sigma\pi}$ 
because $\lambda_{\pi\sigma}<\lambda_{\sigma\pi}$ [28].

\begin{figure}
\begin{center}
\epsfbox{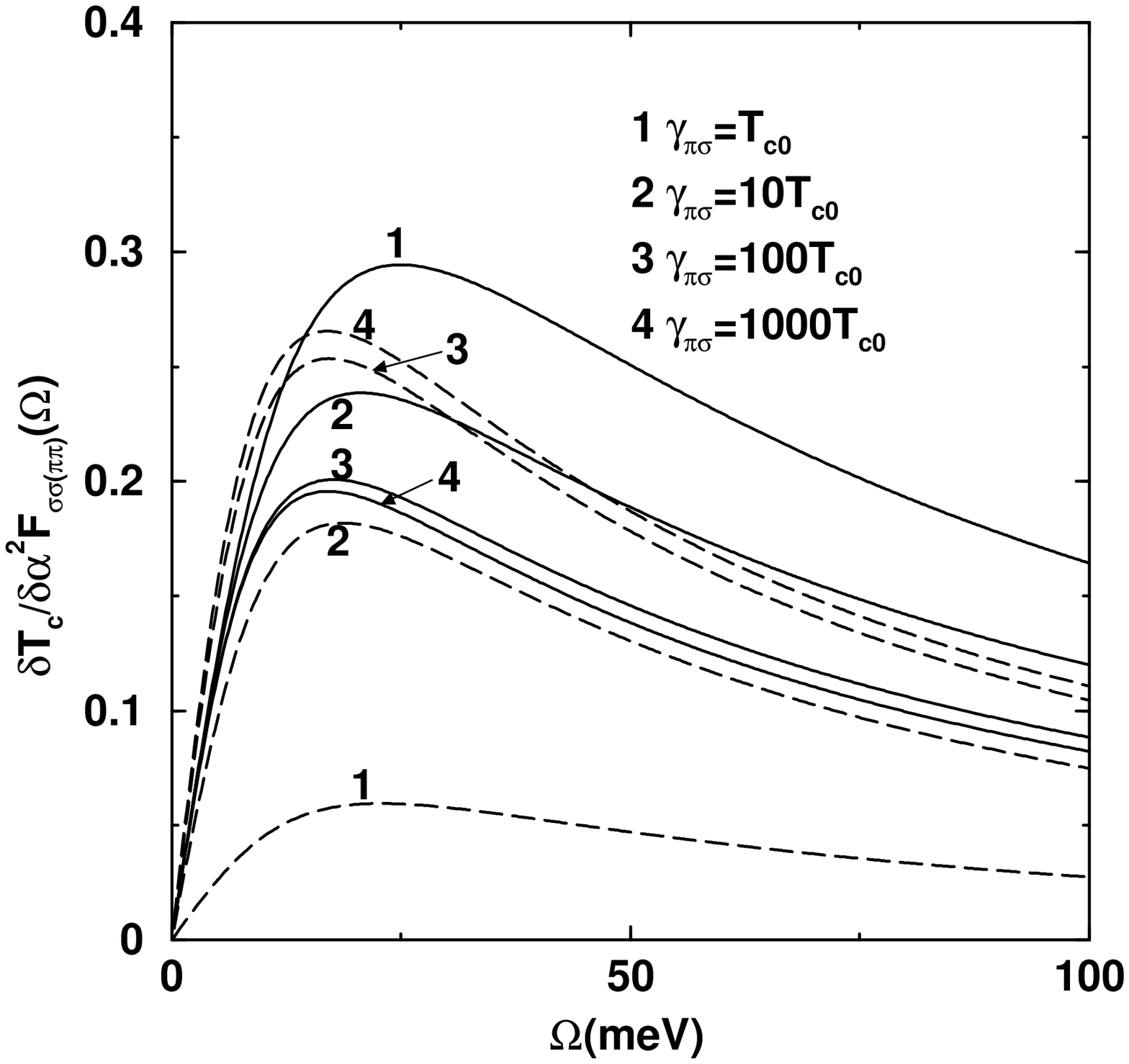}
\end{center}
\caption{\label{label}The band-diagonal functional derivatives $\delta T_{c}/\delta\alpha^2F_{\sigma\sigma}$ 
(solid lines) and $\delta T_{c}/\delta\alpha^2F_{\pi\pi}$ (long dashed lines) for four different 
values of $\gamma_{\pi\sigma}$ given in units of $T_{c0}$.}
\end{figure}

\begin{figure}
\begin{center}
\epsfbox{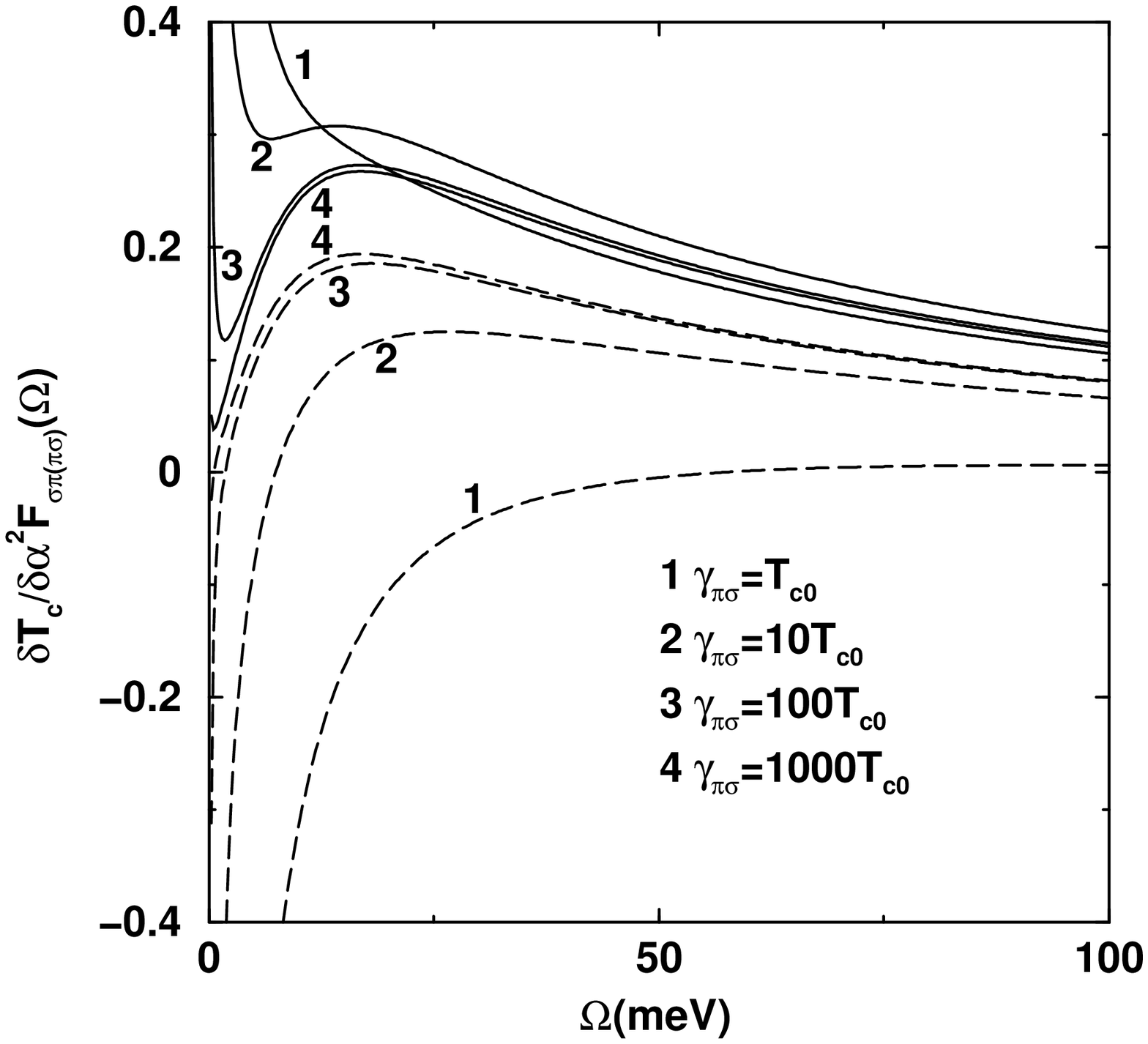}
\end{center}
\caption{\label{label}The band-off-diagonal functional derivatives $\delta T_{c}/\delta\alpha^2F_{\pi\sigma}$
(solid lines) and $\delta T_{c}/\delta\alpha^2F_{\sigma\pi}$ (long dashed lines) for four different
values of $\gamma_{\pi\sigma}$ given in units of $T_{c0}$.}
\end{figure}

\section{$\theta$-$\theta$ model calculation}

The main difference between the strong coupling Eliashberg theory and the BCS approach is that the latter 
does not include the renormalization and the damping effects associated with the electron-phonon interaction. 
In the BCS calculation $Z_{i}^{0}(n)$, Eq.~(4), is set equal to 1. It is possible to improve upon the 
BCS approach so that the effects of renormalization by electron-phonon interaction are included in an 
approximate way through so-called $\theta$-$\theta$ model [26]. In this model $\lambda_{ij}(n-m)$ in 
Eq.~(2) for the electron-phonon contribution to the pairing self-energy is replaced by $\lambda_{ij}\theta(\Omega
_{m}-|\omega_n|)\theta(\Omega_{m}-|\omega_m|)$ with $\Omega_{m}$ - the maximum phonon energy
(BCS approximation) and, after rewriting the sum in (4) as
\[
\sum_{m=-\infty}^{+\infty}\lambda_{ij}(n-m)\frac{\omega_{m}}{|\omega_{m}|} =
2\sum_{m=1}^{n-1}\lambda_{ij}(m) - \lambda_{ij}(0)\>,
\]
$\lambda_{ij}(m)$ in Eq.~(4) for the electron-phonon contribution to the renormalization function 
is replaced by $\lambda_{ij}\theta(\Omega_{m}-|\omega_m|)$. After rescaling the Coulomb repulsion matrix 
from cutoff $\omega_{c}$ to $\Omega_{m}$ according to the general prescription (11), i.e.~$
\hat{\mu}^{*}(\Omega_{m})=[\hat{1}+\hat{\mu}^{*}(\omega_{c})\ln(\omega_{c}/\Omega_{m})]^{-1}
\hat{\mu}^{*}(\omega_{c})$ one gets instead of (7)
\begin{eqnarray}
\fl \Delta_{i}(n)\left(1+\sum_{j}\lambda_{ij}+\sum_{j\neq i}\frac{1}{2\tau_{ij}}\frac{1}{|\omega_{n}|}\right)=
\sum_{j,|\omega_{m}|\leq\Omega_{m}}(\lambda_{ij}-\mu_{ij}^{*}(\Omega_{m}))
\pi T_{c}\frac{\Delta_{j}(m)}{|\omega_{m}|} \nonumber \\
+\sum_{j\neq i}\frac{1}{2\tau_{ij}}\frac{\Delta_{j}(n)}{|\omega_{n}|}\>,
\end{eqnarray}
for $|\omega_{n}|\leq\Omega_{m}$.
After defining
\begin{eqnarray}
\delta_{in} & = & \sqrt{N_{i}}\frac{\Delta_{i}(n)}{|\omega_{n}|}\sqrt{1+\sum_{j}\lambda_{ij}}\>,\\
\Lambda_{ij} & = &\frac{\sqrt{N_{i}/N_{j}}\lambda_{ij}-\mu_{ij}^{*}(\Omega_{m})}{
\sqrt{1+\sum_{k}\lambda_{ik}}\sqrt{1+\sum_{k}\lambda_{jk}}}\>,\\
G_{ij} & = & \frac{1/(2\pi T_{c}t_{ij})}{\sqrt{1+\sum_{k}\lambda_{ik}}\sqrt{1+\sum_{k}\lambda_{jk}}}\>,
\end{eqnarray}
where $t_{11}=\tau_{12}=\tau_{21}N_{1}/N_{2}$, $t_{12}=t_{21}=\tau_{21}\sqrt{N_{1}/N_{2}}$, $t_{22}=\tau_{21}$ 
(in this section we label the $\sigma$ band with 1 and $\pi$ band with 2), Eq.~(16) can be written as  
\begin{equation}
\fl \left[|2n-1|\left( \begin{array}{cc}
 1 & 0\\
 0 & 1
\end{array}\right) + \left( \begin{array}{rr}
G_{11} & -G_{12} \\
-G_{21} & G_{22}
\end{array}\right)\right] \left( \begin{array}{c}
\delta_{1n} \\
\delta_{2n}
\end{array}\right) = \left( \begin{array}{cc}
\Lambda_{11} & \Lambda_{12} \\
\Lambda_{21} & \Lambda_{22}
\end{array}\right) \sum_{|\omega_{m}|\leq\Omega_{m}} \left( \begin{array}{c}
\delta_{1m} \\
\delta_{2m} 
\end{array}\right)
\end{equation}
or, realizing that the right-hand side of (20) does not depend on the Matsubara index and denoting the 
elements of the corresponding 2$\times$1 matrix with $c_{1}$ and $c_{2}$, as 
\begin{equation}
\fl \left( \begin{array}{cc}
\Lambda_{11} & \Lambda_{12} \\
\Lambda_{21} & \Lambda_{22}
\end{array}\right) \sum_{|\omega_{m}|\leq\Omega_{m}}
\left[|2m-1|\left( \begin{array}{cc}
1 & 0\\
 0 & 1
\end{array}\right)+
\left( \begin{array}{rr}
G_{11} & -G_{12} \\
-G_{21} & G_{22}
\end{array}\right)\right]^{-1} \left( \begin{array}{c}
c_1 \\
c_2
\end{array}\right) =  \left( \begin{array}{c}
c_1 \\
c_2
\end{array}\right).
\end{equation}
The 2$\times$2 matrix 
\begin{equation}
\hat{G}=\left( \begin{array}{rr}
G_{11} & -G_{12} \\
-G_{21} & G_{22}
\end{array}\right)
\end{equation}
is a real symmetric matrix, Eq.~(19), with eigenvalues $d=G_{11}+G_{22}$ and 0 (see (19) and subsequent 
definitions of various $t_{ij}$ parameters)
and could be diagonalized through an orthogonal transformation 
\begin{equation}
\hat{R}\hat{G}\hat{R}^{-1}= \left( \begin{array}{cc}
d & 0 \\
0 & 0
\end{array}\right)\>,
\end{equation}
where the elements of $\hat{R}$ are $R_{11}=\sqrt{G_{11}/(G_{11}+G_{22})}$, 
$R_{12}=-\sqrt{G_{22}/(G_{11}+G_{22})}$, $R_{21}=-R_{12}$ and $R_{22}=R_{11}$. Expressing $\hat{G}$ in (21) 
in terms of the right-hand side of Eq.~(23) and using 
\begin{equation}
\sum_{|\omega_{m}|\leq\Omega_{m}}\frac{1}{|2m-1|+d}=\psi\left(\frac{\Omega_{m}}{2\pi T_{c}}+1+\frac{d}{2}\right)
-\psi\left(\frac{1}{2}+\frac{d}{2}\right)\>,
\end{equation}
where $\psi$ is the digamma function [26], Eq.~(21) can be rewritten as
\begin{equation}
\hat{M}\left( \begin{array}{c}
c_1 \\
c_2
\end{array}\right) = \left( \begin{array}{c}
c_1 \\
c_2
\end{array}\right)\>,
\end{equation}
where
\begin{eqnarray}
\fl \hat{M} = \left[\psi\left(\frac{\Omega_{m}}{2\pi T_{c}}+1\right)-\psi\left(\frac{1}{2}\right)\right]
\left( \begin{array}{cc}
\Lambda_{11} & \Lambda_{12} \\
\Lambda_{21} & \Lambda_{22}
\end{array}\right) - 
\left[\psi\left(\frac{G_{11}+G_{22}}{2}+\frac{1}{2}\right)
-\psi\left(\frac{1}{2}\right)\right. \nonumber \\
 \left. +\psi\left(\frac{\Omega_{m}}{2\pi T_{c}}+1\right)-\psi\left(\frac{\Omega_{m}}{2\pi T_{c}}+1+ 
\frac{G_{11}+G_{22}}{2}\right)\right] \nonumber \\
\times\left( \begin{array}{cc}
\Lambda_{11} & \Lambda_{12} \\
\Lambda_{21} & \Lambda_{22}
\end{array}\right)
\left( \begin{array}{cc}
\frac{G_{11}}{G_{11}+G_{22}} & \frac{-\sqrt{G_{11}G_{22}}}{G_{11}+G_{22}} \\
\frac{-\sqrt{G_{11}G_{22}}}{G_{11}+G_{22}} & \frac{G_{22}}{G_{11}+G_{22}}
\end{array}\right)\>.
\end{eqnarray}
The transition temperature is the highest $T_{c}$ for which the larger eigenvalue of $\hat{M}$ is 
equal to 1.

We have solved Eqs.~(25-26) for $T_{c}$ as a function of the interband impurity scattering rate 
and our results are shown by the long dashed line in Fig.~1. 
The electron-phonon interaction parameters were taken to be the same as those used in section 2: 
$\lambda_{11}\equiv 
\lambda_{\sigma\sigma}$ = 1.017, $\lambda_{12}\equiv\lambda_{\sigma\pi}$ = 0.212, $\lambda_{22}\equiv
\lambda_{\pi\pi}$ = 0.446 and $\lambda_{21}\equiv\lambda_{\pi\sigma}$ = 0.155. The maximum phonon 
energy was taken to be $\Omega_{m}$ = 75 meV, which is roughly the position of the largest peak in 
$\alpha^{2}F_{\sigma\sigma}$ (see Fig.~1 in Ref.~[25]) and the values of $\mu^{*}(\omega_{c})$s 
from section 2.2 were scaled down using (11) to the new cutoff $\Omega_{m}$ to give
$\mu_{11}^{*}(\Omega_{m})\equiv\mu_{\sigma\sigma}^{*}(\Omega_{m})$ = 0.139578, 
$\mu_{22}^{*}(\Omega_{m})\equiv\mu_{\pi\pi}^{*}(\Omega_{m})$ = 0.139217, 
$\mu_{12}^{*}(\Omega_{m})=\mu_{21}^{*}(\Omega_{m})$ = 0.027081. Clearly, including the electron-phonon 
renormalization effects improves the BCS treatment considerably. However, we want to stress 
that $\theta$-$\theta$ model gives the improved values {\em only} for the {\em reduced} quantity 
$T_{c}/T_{c0}$ as a function of the {\em reduced} interband scattering rate $\gamma_{\pi\sigma}/T_{c0}$.  
The {\em absolute} values of $T_{c}$ are not accurately predicted by $\theta$-$\theta$ model (e.g. we 
get too large a value for $T_{c0}$ of 143 K so that the usual weak coupling approximation 
$\psi(\Omega_{m}/(2\pi T_{c0})+1)-\psi(1/2)\approx\ln(2e^{\gamma}\Omega_{m}/(\pi T_{c0}))$, 
where $\gamma$ is the Euler's constant, cannot be made).

\section{Summary}

We have calculated the change in the superconducing transition temperature of MgB$_{2}$ caused by 
interband nonmagnetic impurity scattering using the Eliashberg theory with realistic 
electron-phonon [16] and Coulomb repulsion [24] parameters for this compound. Our central result 
is given in Fig.~1. We find much slower rate of $T_{c}$ suppression than what is obtained from 
the BCS approach [19] which ignores the renormalization and damping effects associated with the 
electron-phonon interactions. For small interband scattering rates the strong coupling calculation 
gives about 4.5 times slower suppression rate of $T_{c}$ than the BCS approach. Moreover, the 
strong coupling calculation indicates that it is unrealistic to expect the transition 
temperature of MgB$_{2}$ to ever drop below 60\% of its maximum value as a result of 
impurity scattering and the 20\% drop in $T_{c}$ upon irradiation by fast neutrons [13] is 
certainly within our calculated range (20\% drop in $T_{c}$ is obtained with $\gamma_{\pi\sigma}$ 
of about 4$k_{B}T_{c0}$, Fig.~1).
Hence,
the initial expectations based on the BCS treatment of the two-band model [19] that a dramatic suppression 
of $T_{c}$ in MgB$_{2}$ with interband impurity scattering would provide the final ``smoking gun'' 
evidence for the two-band model was exaggerated. 

Our calculation with $\theta$-$\theta$ model (long dashed line in Fig.~1) clearly indicates that the 
main reason for the failure of the BCS approach to quantitatively account for the dependence of 
$T_{c}/T_{c0}$ on $\gamma_{\pi\sigma}/T_{c0}$ in MgB$_{2}$
is that the BCS treatment leaves out the electron-phonon renormalization effects. One should 
keep in mind, however, that for other multiband systems with electron-phonon and 
Coulomb interactions different from those calculated [16,24] for MgB$_{2}$ one would have to 
recalculate $T_{c}/T_{c0}$ as a function of interband scattering rate $\gamma_{\pi\sigma}/T_{c0}$ 
using Eliashberg equations from section 2.1 with the interaction parameters which are relevant 
to the multiband superconductor that is being considered. 
\ack
This work has been supported in part by the Natural Sciences and Engineering 
Research Council of Canada. The author is grateful to O.~Jepsen for 
providing the numerical values of $\alpha^{2}F$s for MgB$_{2}$ presented in [16] and to K.~V.~Samokhin, 
M.~Reedyk and S.~K.~Bose for their interest in this work.
\end{document}